\documentclass[prb,twocolumn,preprintnumbers,amsmath,amssymb]{revtex4}

\usepackage{graphicx}% Include figure files
\usepackage{dcolumn}% Align table columns on decimal point
\usepackage{bm}% bold math

\begin{document}

%\preprint{APS/123-QED}

\title{Room temperature Optical Orientation of Exciton Spin \\in cubic GaN/AlN quantum dots}

\author{D. Lagarde, A. Balocchi, H. Carr\`ere, P. Renucci, T. Amand, X. Marie}
 \email{marie@insa-toulouse.fr}
\affiliation{Laboratoire de Physique et Chimie des Nano-Objets, INSA-CNRS-UPS, 135 Avenue de Rangueil, 31077 Toulouse Cedex 4, France
}

\author{S. Founta, H. Mariette}
\affiliation{
CEA-CNRS group \textquotedblleft Nanophysique et Semiconducteurs\textquotedblright , Institut N\'eel-CNRS and CEA/DRFMC/SP2M, 25 Rue des martyrs, 38042 Grenoble, France}%

\date{\today}

\begin{abstract}
The optical orientation of the exciton spin in an ensemble of self-organized cubic GaN/AlN quantum dots is studied by time-resolved photoluminescence. Under a polarized quasi-resonant excitation, the luminescence linear polarization exhibits no temporal decay, even at room temperature. This demonstrates the robustness of the exciton spin polarization in these cubic nitride nanostructures, with characteristic decay times longer than 10 ns.
\end{abstract}
\maketitle
Wide bandgap GaN-based semiconductors and related heterostructures represent promising candidates for spintronic applications\cite{Beschoten, Ohno, Buyanova}. The weak spin-orbit coupling in these materials should yield long electron spin relaxation times \cite{Krishnamurthy_2003} and their large exciton binding energy ($\sim$26 meV in bulk GaN) should allow the control and manipulation of the exciton spin, even at high temperature. \\
Compared to GaAs-based structures, very few measurements of the carrier spin properties have been performed on these wide bandgap nitrides which can crystallize either in the wurtzite (Wz) or zinc-blende cubic (ZB) structure \cite{Paisley}. Most of these studies deal with the measurement of the electron or hole spin dynamics in Wz GaN or InGaN structures. Beschoten \textit{et al.} measured an \textit{electron} spin lifetime of $\sim$35 ps at room temperature in \textit{n}-doped Wz epilayers \cite{Beschoten}. A \textit{hole} spin coherence time of $\sim$120 ps was found in \textit{p}-type bulk Wz GaN at low temperature \cite{Ohno}. However, a very fast \textit{exciton} spin relaxation time, of the order of 1 ps, was deduced from transient reflectivity or spin grating experiments in non-intentionally doped Wz GaN epilayers \cite{Tackeuchi_2004, Ishiguro}. In Wz nitride nanostructures (quantum wells or quantum dots), \textit{exciton} spin relaxation times of the order of 200 ps were also reported at $T=300$ K with a rather weak temperature dependence \cite{Nagahara_2005, Nagahara_2006}. In all these wurtzite nitride structures, the electronic and spin properties are highly affected by the strong built-in electric field due to the spontaneous and piezoelectric polarizations \cite{Fonoberov, Julier}. \\
In contrast, these polarizations are negligible in zinc-blende GaN structures, due to the high crystal symmetry and much longer carrier spin relaxation times can be expected. Krishnamurthy \textit{et al.} calculated that the electron spin lifetime in pure ZB GaN is about three orders of magnitude larger than in GaAs at all temperatures as a result of the lower spin-orbit interaction and higher conduction band density of states \cite{Krishnamurthy_2003}. An electron spin lifetime as long as 100 ns is predicted at $T=300$ K in high quality bulk GaN, dropping significantly if the defect density increases. In a ZB GaN epilayer, Tackeuchi \textit{et al.} measured indeed a much longer bound exciton spin relaxation time ($\sim$5 ns at 15 K) compared to Wz structures but surprisingly no spin memory was found above 75 K \cite{Tackeuchi_2006}. No report has been published to date on the measurement of the spin dynamics in ZB nitride nanostructures, in which the influence of defects on the electronic properties is usually weaker thanks to the carrier confinement. 

We present in this letter a detailed time-resolved investigation of the exciton spin dynamics in self-organized ZB GaN/AlN quantum dots (QDs). These optical orientation experiments clearly evidence an exciton linear polarization which persists at room temperature, in contrast to the ususal behaviour in other III-V or II-VI nanostructures \cite{Paillard, Kalt, Henneberger}. 

The QD sample investigated here consists in 18 self-assembled ZB GaN quantum dot planes in AlN barriers, grown on 3C-SiC(001)/Si pseudosubstrate by molecular-beam epitaxy \cite{Mariette}. The dots have a typical diameter of 12 nm and a height around 1.5 nm. The sample is nominally undoped and the areal dot density is of the order of $1\cdot10^{11}$ cm$^{-2}$. In time-resolved photoluminescence (PL) experiments, the excitation source is provided by a mode-locked frequency tripled Ti:Sa laser, with a 1.5 ps pulse width and a tunable wavelength in the range 260-310 nm. The laser beam is focused onto the sample to a 100 $\mu$m diameter spot with an average power $P_{exc}=1$ mW \cite{Puissance}. The PL signal is dispersed by an imaging spectrometer and then temporally resolved by a S20 photocathode streak camera with an overall time resolution of 8 ps. The linear polarization degree of the luminescence is defined as $P_{Lin} = (I^X - I^Y) / (I^X + I^Y)$. Here $I^X (I^Y)$ denotes the $X (Y)$ linearly polarized PL components ($X$ and $Y$ are chosen parallel to the [110] and [1$\overline{1}$0] crystallographic directions). 
\begin{figure}
\includegraphics[width=0.5\textwidth]{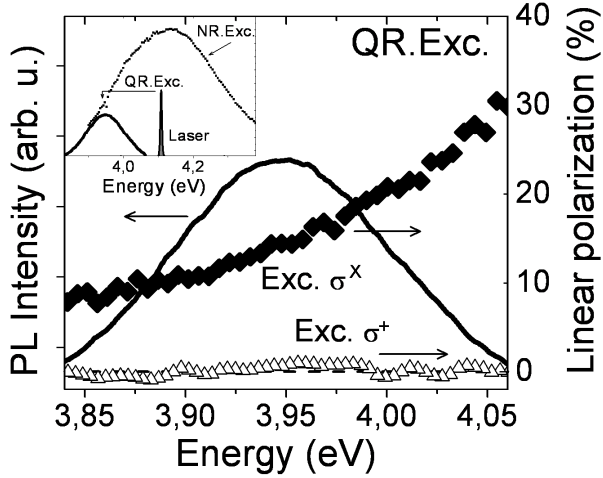}
\caption{Time-integrated PL spectrum under a quasi-resonant excitation at $T=20$ K. The linear polarization degree following a linearly-polarized ($\blacklozenge$) or circularly-polarized ($\triangle$) excitation is also plotted. Inset: Time-integrated PL spectra of the QD for quasi-resonant excitation ($E_{exc}=4.11$ eV) and non-resonant excitation ($E_{exc}=4.77$ eV).}
\label{fig1}
\end{figure}

The inset of Fig.~\ref{fig1} presents the time-integrated PL spectrum measured at $T=20$ K for a non-resonant excitation energy $E_{exc} = 4.77$ eV. The ground state emission is centered around 4.15 eV with a full width at half maximum of 250 meV which results from the inhomogeneous broadening of the ground state exciton energies due to the size and strain distributions among the QD ensemble. The PL spectrum obtained under Quasi-Resonant (QR) excitation ($E_{exc}=4.11$ eV) is also plotted. QR excitation means here that the laser excitation energy lies within the QD ground state energy range. A much narrower PL emission spectrum is measured in this case due to the spectral selectivity of the excitation \cite{Heitz}. The linear polarization of the time-integrated\cite{Integration} exciton PL emission following a linearly-polarized ($\sigma^X$) QR excitation is presented in Fig.~\ref{fig1}. A clear linear polarization is observed for any detection energy in the spectrum, with a larger value ($P_{Lin}\sim 30$ $\%$) on the high energy part. This linear polarization arises from the optical orientation of the exciton spin (also called exciton optical alignment), as already observed in many III-V or II-VI bulk materials or heterostructures \cite{Planel, Orientation, Paillard, Kalt}. The circular or linear exciton polarization dynamics can be described in the framework of an effective pseudospin with $S=1/2$ \cite{Ivchenko}. In this formalism, the exciton states with angular momentum $\left | +1 \right \rangle$ and $\left | -1 \right \rangle$ are equivalent to a pseudospin polarized parallel or anti-parallel to the \textit{z}-axis ($S_z=+\frac{1}{2}$ or $-\frac{1}{2} $). The quantization axis ($Oz$) is chosen along the light propagation direction which is also the sample growth direction. The linear exciton states $\left | X \right \rangle$ and $\left | Y \right \rangle$ are described by a pseudo-spin $S_x=+\frac{1}{2}$ or $-\frac{1}{2} $, respectively. Under a linearly polarized excitation ($\sigma^X$), the initial exciton pseudospin $\overrightarrow{S}(0)$ is parallel to the ($Ox$) axis. The linear exciton PL polarization writes simply $P_{Lin}=2S_x$. The absence of linear polarization following a circularly-polarized excitation ($\sigma^+$), as shown in Fig.~\ref{fig1}, (open triangles) proves that the measured polarization (solid diamonds) does not arise from the valence state mixing induced by the strain or by the QD shape anisotropy, as it was observed in Wz InGaN QDs \cite{Bimberg, Angle}. Indeed, such effect would yield a linearly-polarized exciton emission whatever the polarization of the excitation light is. The absence of linear polarization when the laser excitation energy increases (as detailed below) is another confirmation that the observed linear polarization in Fig.~\ref{fig1} is indeed due to the light-induced optical alignment of excitons in the QDs. Let us furthermore recall that the observation of optical pumping in linear configuration is a clear signature of intrinsic exciton emission \cite{Planel} : the loss of coherence between electron and hole spins, inherent in the impurity bound excitons or charged excitons, inhibits the optical alignment of these pseudo-particles.
\begin{figure}
\includegraphics[width=0.5\textwidth]{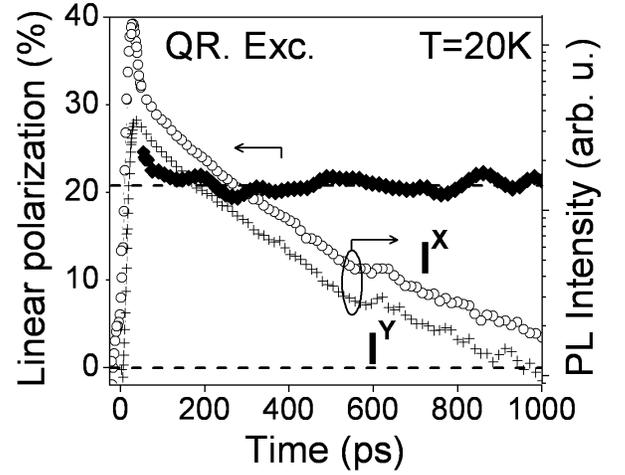}
\caption{Temporal dependence of the PL components co-polarized $I^X$ ($\circ$) and counter-polarized $I^Y$ (+) to the $\sigma^X$ polarized excitation laser at $T=20$ K and the corresponding linear polarization degree $P_{Lin}$ ($\blacklozenge$). $E_{exc}=4.11$ eV, $E_{det}=4.02$ eV.} 
\label{fig2}
\end{figure}

Fig. \ref{fig2} displays the time evolution of the co-polarized $I^X$ and counter-polarized $I^Y$ PL intensity components obtained after a linearly-polarized $\sigma^X$ excitation (the detection energy is $E_{det}=4.02eV$). The decay time of the PL intensity is about 350 ps, much shorter than in Wz GaN structures because of the absence of the internal fields \cite{Simon,Grandjean}. 
It is clear from Fig.~\ref{fig2} that the QD emission exhibits a linear polarization $P_{Lin}\sim 20$ $\%$ which remains strictly constant in time within our experimental accuracy during the exciton emission. This behaviour is the same whatever the detection energy is within the PL spectrum (not shown). Using an exponential fitting procedure, we can infer that the linear polarization decay time is longer than 10 ns. This result differs strongly from the exciton polarization dynamics in Wz-type structures, characterized by a polarization decay time 2 or 3 orders of magnitude shorter \cite{Tackeuchi_2004, Nagahara_2006}.

The excitonic properties of nitride-based ZB QDs are still poorly understood. However, from similar results on InAs, CdTe or CdSe self-organized QDs \cite{Paillard, Kalt, Henneberger}, the experimental data presented above strongly suggest that the eigenstates of the exciton in the GaN QDs are linearly-polarized states, aligned along the [110] and [1$\overline{1}$0] directions. This exciton fine structure may originate from QD elongation and/or interface anisotropy which yield a significant anisotropic exchange interaction between the electron and the hole forming the exciton. No circular polarization of the PL emission was observed following a circularly-polarized excitation. Similarly, no PL linear polarization was detected after a linearly-polarized excitation along [100] or [010] (not shown). This is consistent with $\left | X \right \rangle$ and $\left | Y \right \rangle$ linearly-polarized exciton eigenstates in an inhomogeneous QD ensemble. When the polarized excitation creates a coherent superposition of these exciton eigenstates, the anisotropic exchange energy statistical fluctuations among the detected QD ensemble prevent the observation of spin quantum beats~\cite{Paillard}.
\begin{figure}
\includegraphics[width=0.5\textwidth]{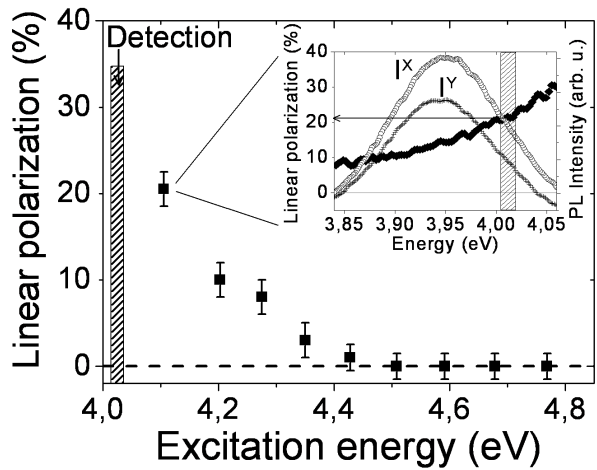}
\caption{PL linear polarization detected at a fixed energy $E_{det}=4.02$ eV as a function of excitation energy at $T=20$ K. Inset: Spectral dependence of the co- ($I^X$) and counter-polarized ($I^Y$) PL components and the associated linear polarization degree for $E_{exc}=4.11$ eV.}
\label{fig3}
\end{figure}
\begin{figure}
\includegraphics[width=0.5\textwidth]{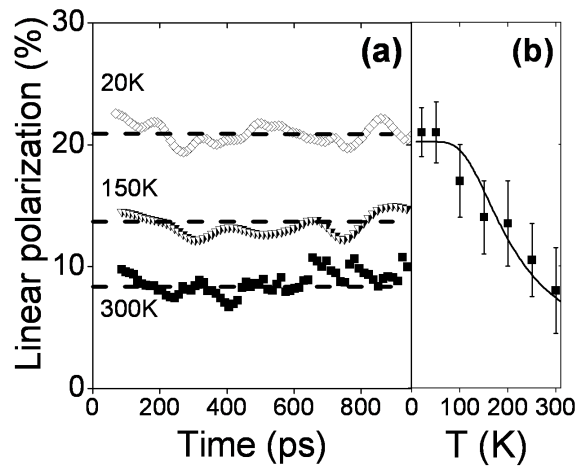}
\caption{(a) Linear polarization dynamics for $T=20$, 150 and 300 K. (b) Temperature dependence of the linear polarization degree (the full line is a guide to the eyes). For $T=20$ K, the excitation and detection energies are $E_{exc}=4.11$ eV and $E_{det}=4.02$ eV, respectively. For higher temperature, the excitation and detection conditions are detailed in the text.}
\label{fig4}
\end{figure}

The relative low value of the linear polarization ($P_{Lin}\sim$20 $\%$) observed in Fig.~\ref{fig2} is mainly due to the fact that the experiments are not performed under strictly resonant conditions (which could not be performed with our present experimental set-up because of the large laser scattering on the sample surface). In these QR excitation conditions, both the energy relaxation processes and the excitation of exciton states with different symmetries contribute to decrease the measured $P_{Lin}$. Nevertheless, the exciton spin relaxation quenching, evidenced by the absence of decay of the linear polarization, is a striking feature here. Already observed in other self-organized QD structures at low temperature, it proves that neither the electron nor the hole spin relax on the exciton lifetime scale and that once the exciton occupies the ground state of the QD, no transient change of the polarization occurs within the time window under investigation \cite{Paillard, Steel, Henneberger}.

Fig.~\ref{fig3} displays the degree of the PL linear polarization detected at a fixed energy ($E_{det}=4.01$ eV) as a function of the excitation energy. The average $P_{Lin}$ decreases from 20 to 0 $\%$ when the excess energy $E_{exc}-E_{det}$ increases from $\sim$90 to 400 meV. This behavior is consistent with the variation of the linear polarization in the PL spectrum measured in Fig.~\ref{fig1}. The decrease of $P_{Lin}$ appears as a monotonic function of the excess energy, in agreement with the pioneering work in bulk CdS or recent results in CdSe/ZnSe self-organized QDs \cite{Planel, Kusrayev}. Nevertheless, no distinct peaks due to LO-phonon cascade processes are evidenced. The absence of such peaks in Fig.~\ref{fig3} is probably due to the large inhomogeneous broadening in ZB GaN QDs. We emphasize that for $E_{exc}-E_{det}>$ 400 meV, no linear polarization is measured for any detection energy in the PL spectrum. 

Fig~\ref{fig4}(a) presents the dependence of the exciton PL linear polarization dynamics upon the lattice temperature. As the amplitude of the linear polarization depends on the detection energy in the spectrum (Fig.~\ref{fig1}), we varied the excitation and detection energies for the different temperatures following a simple Varshni law \cite{Varshni} keeping constant $E_{exc}-E_{det}=90$ meV. This makes possible the measurement of the temperature dependence of the exciton spin properties for the same QDs. It is worth mentioning that we obtained qualitatively the same results for fixed excitation and detection energies (not shown). Contrary to the reported results in ZB bulk GaN \cite{Tackeuchi_2006}, optical orientation of exciton spin is clearly observed in Fig.~\ref{fig4} above 75 K. Moreover, no temporal decay of $P_{Lin}$ is observed, even for high temperatures. This result is to be compared with those obtained on QDs made of other semiconductor materials (InAs, CdSe, CdTe)\cite{Paillard, Kalt, Henneberger}. In these QDs, the linear polarization decay time, usually longer than the exciton lifetime at T=10 K, drastically drops when the temperature increases a few tens of K \cite{Paillard, Kalt}. Thus, the robustness of the exciton linear polarization up to room temperature appears here as a unique property of the ZB GaN QDs. Though we have never observed any temporal decay of $P_{Lin}$ on the exciton lifetime scale, a clear decrease of the amplitude of the linear polarization is evidenced. The uncertainty on the measurements in Fig.~\ref{fig4}(b) does not allow us to extract an accurate thermal activation energy (it lies within the 50-100 meV range which is a plausible value for the exciton binding energy in these structures \cite{Ramvall, Xia}).

Finally, we have investigated the possible conversion from optical alignment to optical orientation of excitons when a magnetic field is applied along the growth direction (Faraday geometry) \cite{Dzhioev, Paillard}. If the exciton Zeeman splitting $\hbar\Omega_z=g\mu_BB$ is much larger than the anisotropic exchange energy, the QD exciton eigenstates are no longer the exciton $\left | X \right \rangle$ and $\left | Y \right \rangle$ linearly-polarized states but the $\left | +1 \right \rangle$ and $\left | -1 \right \rangle$ circular ones. However, this conversion has not been observed for any magnetic field value up to 4 T (the maximum field available on our experimental set-up), probably because of the very large value of the exchange energy compared to the exciton Zeeman term. Much stronger magnetic fields would be required to evidence these effects.

In conclusion, we have studied the optical orientation of exciton spin in self-organized cubic GaN/AlN quantum dots. We observe no measurable temporal decay of the exciton linear luminescence polarization at any investigated temperature. Even at room temperature, the exciton spin decay time is larger than 10 ns, i.e. 2 or 3 orders of magnitude longer than in wurtzite-type structures. Moreover, these results contrast with the fast exciton spin relaxation previously observed in other III-V or II-VI QD systems at high temperature. 

The authors are grateful to EADS Research Foundation and to Institut Universitaire de France for financial support. They are also grateful to the NOVASiC compagny which provides them with the 3C-SiC(001)/Si pseudosubstrates and with the financial support of S.F. PhD fellowship. They thank B. Daudin and B. Gayral (Grenoble) for fruitful discussions. 

\begin{thebibliography}{32}
\expandafter\ifx\csname natexlab\endcsname\relax\def\natexlab#1{#1}\fi
\expandafter\ifx\csname bibnamefont\endcsname\relax
  \def\bibnamefont#1{#1}\fi
\expandafter\ifx\csname bibfnamefont\endcsname\relax
  \def\bibfnamefont#1{#1}\fi
\expandafter\ifx\csname citenamefont\endcsname\relax
  \def\citenamefont#1{#1}\fi
\expandafter\ifx\csname url\endcsname\relax
  \def\url#1{\texttt{#1}}\fi
\expandafter\ifx\csname urlprefix\endcsname\relax\def\urlprefix{URL }\fi
\providecommand{\bibinfo}[2]{#2}
\providecommand{\eprint}[2][]{\url{#2}}

\bibitem[{\citenamefont{Beschoten et~al.}(2001)\citenamefont{Beschoten,
  Johnston-Halperin, Young, Poggio, J.~E.~Grimaldi, Keller, DenBaars, Mishra, Hu, and
  Awschalom}}]{Beschoten}
\bibinfo{author}{\bibfnamefont{B.}~\bibnamefont{Beschoten}},
  \bibinfo{author}{\bibfnamefont{E.}~\bibnamefont{Johnston-Halperin}},
  \bibinfo{author}{\bibfnamefont{D.~K.} \bibnamefont{Young}},
  \bibinfo{author}{\bibfnamefont{M.}~\bibnamefont{Poggio}},
  \bibinfo{author}{\bibfnamefont{J.~E.} \bibnamefont{Grimaldi}},
  \bibinfo{author}{\bibfnamefont{S.} \bibnamefont{Keller}},
  \bibinfo{author}{\bibfnamefont{S.~P.} \bibnamefont{DenBaars}},
  \bibinfo{author}{\bibfnamefont{U.~K.} \bibnamefont{Mishra}},
  \bibinfo{author}{\bibfnamefont{E.~L.} \bibnamefont{Hu}}, \bibnamefont{and}
  \bibinfo{author}{\bibfnamefont{D.~D.} \bibnamefont{Awschalom}},
  \bibinfo{journal}{Phys. Rev. B} \textbf{\bibinfo{volume}{63}},
  \bibinfo{pages}{121202(R)} (\bibinfo{year}{2001}).

\bibitem[{\citenamefont{Hu et~al.}(2005)\citenamefont{Hu, Morita, Sanada,
  Matsuzaka, Ohno, and Ohno}}]{Ohno}
\bibinfo{author}{\bibfnamefont{C.~Y.}~\bibnamefont{Hu}},
  \bibinfo{author}{\bibfnamefont{K.}~\bibnamefont{Morita}},
  \bibinfo{author}{\bibfnamefont{H.}~\bibnamefont{Sanada}},
  \bibinfo{author}{\bibfnamefont{S.}~\bibnamefont{Matsuzaka}},
  \bibinfo{author}{\bibfnamefont{Y.}~\bibnamefont{Ohno}}, \bibnamefont{and}
  \bibinfo{author}{\bibfnamefont{H.}~\bibnamefont{Ohno}},
  \bibinfo{journal}{Phys. Rev. B} \textbf{\bibinfo{volume}{72}},
  \bibinfo{pages}{121203(R)} (\bibinfo{year}{2005}).

\bibitem[{\citenamefont{Chen et~al.}(2005)\citenamefont{Chen, Buyanova,
  Nishibayashi, Kayanuma, Seo, Murayama, Oka, Thaler, Frazier, Abernathy
  et~al.}}]{Buyanova}
\bibinfo{author}{\bibfnamefont{W.}~\bibnamefont{Chen}},
  \bibinfo{author}{\bibfnamefont{I.}~\bibnamefont{Buyanova}},
  \bibinfo{author}{\bibfnamefont{K.}~\bibnamefont{Nishibayashi}},
  \bibinfo{author}{\bibfnamefont{K.}~\bibnamefont{Kayanuma}},
  \bibinfo{author}{\bibfnamefont{K.}~\bibnamefont{Seo}},
  \bibinfo{author}{\bibfnamefont{A.}~\bibnamefont{Murayama}},
  \bibinfo{author}{\bibfnamefont{Y.}~\bibnamefont{Oka}},
  \bibinfo{author}{\bibfnamefont{G.}~\bibnamefont{Thaler}},
  \bibinfo{author}{\bibfnamefont{R.}~\bibnamefont{Frazier}},
  \bibinfo{author}{\bibfnamefont{C.}~\bibnamefont{Abernathy}},
  \bibnamefont{et~al.}, \bibinfo{journal}{Appl. Phys. Lett.}
  \textbf{\bibinfo{volume}{87}}, \bibinfo{pages}{192107}
  (\bibinfo{year}{2005}).

\bibitem[{\citenamefont{Krishnamurthy et~al.}(2003)\citenamefont{Krishnamurthy,
  Schilfgaarde, and Newman}}]{Krishnamurthy_2003}
\bibinfo{author}{\bibfnamefont{S.}~\bibnamefont{Krishnamurthy}},
  \bibinfo{author}{\bibfnamefont{M.~V.} \bibnamefont{Schilfgaarde}},
  \bibnamefont{and} \bibinfo{author}{\bibfnamefont{N.}~\bibnamefont{Newman}},
  \bibinfo{journal}{Appl. Phys. Lett.} \textbf{\bibinfo{volume}{83}},
  \bibinfo{pages}{1761} (\bibinfo{year}{2003}).

\bibitem[{\citenamefont{Paisley et~al.}(1989)\citenamefont{Paisley, Sitar,
  Posthill, and Davis}}]{Paisley}
\bibinfo{author}{\bibfnamefont{M.}~\bibnamefont{Paisley}},
  \bibinfo{author}{\bibfnamefont{Z.}~\bibnamefont{Sitar}},
  \bibinfo{author}{\bibfnamefont{J.}~\bibnamefont{Posthill}}, \bibnamefont{and}
  \bibinfo{author}{\bibfnamefont{R.}~\bibnamefont{Davis}}, \bibinfo{journal}{J.
  Vac. Sci. Technol. A} \textbf{\bibinfo{volume}{7}}, \bibinfo{pages}{701}
  (\bibinfo{year}{1989}).

\bibitem[{\citenamefont{Kuroda et~al.}(2004)\citenamefont{Kuroda, Yabushita,
  Kosuge, Tackeuchi, Taniguchi, Chinone, and Horio}}]{Tackeuchi_2004}
\bibinfo{author}{\bibfnamefont{T.}~\bibnamefont{Kuroda}},
  \bibinfo{author}{\bibfnamefont{T.}~\bibnamefont{Yabushita}},
  \bibinfo{author}{\bibfnamefont{T.}~\bibnamefont{Kosuge}},
  \bibinfo{author}{\bibfnamefont{A.}~\bibnamefont{Tackeuchi}},
  \bibinfo{author}{\bibfnamefont{K.}~\bibnamefont{Taniguchi}},
  \bibinfo{author}{\bibfnamefont{T.}~\bibnamefont{Chinone}}, \bibnamefont{and}
  \bibinfo{author}{\bibfnamefont{N.}~\bibnamefont{Horio}},
  \bibinfo{journal}{Appl. Phys. Lett.} \textbf{\bibinfo{volume}{85}},
  \bibinfo{pages}{3116} (\bibinfo{year}{2004}).

\bibitem[{\citenamefont{Ishiguro et~al.}(2007)\citenamefont{Ishiguro, Toda, and
  Adachi}}]{Ishiguro}
\bibinfo{author}{\bibfnamefont{T.}~\bibnamefont{Ishiguro}},
  \bibinfo{author}{\bibfnamefont{Y.}~\bibnamefont{Toda}}, \bibnamefont{and}
  \bibinfo{author}{\bibfnamefont{S.}~\bibnamefont{Adachi}},
  \bibinfo{journal}{Appl. Phys. Lett.} \textbf{\bibinfo{volume}{90}},
  \bibinfo{pages}{011904} (\bibinfo{year}{2007}).

\bibitem[{\citenamefont{Nagahara et~al.}(2005)\citenamefont{Nagahara, Arita,
  and Arakawa}}]{Nagahara_2005}
\bibinfo{author}{\bibfnamefont{S.}~\bibnamefont{Nagahara}},
  \bibinfo{author}{\bibfnamefont{M.}~\bibnamefont{Arita}}, \bibnamefont{and}
  \bibinfo{author}{\bibfnamefont{Y.}~\bibnamefont{Arakawa}},
  \bibinfo{journal}{Appl. Phys. Lett.} \textbf{\bibinfo{volume}{86}},
  \bibinfo{pages}{242103} (\bibinfo{year}{2005}).

\bibitem[{\citenamefont{Nagahara et~al.}(2006)\citenamefont{Nagahara, Arita,
  and Arakawa}}]{Nagahara_2006}
\bibinfo{author}{\bibfnamefont{S.}~\bibnamefont{Nagahara}},
  \bibinfo{author}{\bibfnamefont{M.}~\bibnamefont{Arita}}, \bibnamefont{and}
  \bibinfo{author}{\bibfnamefont{Y.}~\bibnamefont{Arakawa}},
  \bibinfo{journal}{Appl. Phys. Lett.} \textbf{\bibinfo{volume}{88}},
  \bibinfo{pages}{083101} (\bibinfo{year}{2006}).

\bibitem[{\citenamefont{Fonoberov and Balandin}(2003)}]{Fonoberov}
\bibinfo{author}{\bibfnamefont{V.~A.} \bibnamefont{Fonoberov}}
  \bibnamefont{and} \bibinfo{author}{\bibfnamefont{A.~A.}
  \bibnamefont{Balandin}}, \bibinfo{journal}{J. Appl. Phys.}
  \textbf{\bibinfo{volume}{94}}, \bibinfo{pages}{7178} (\bibinfo{year}{2003}).

\bibitem[{\citenamefont{Julier et~al.}(1999)\citenamefont{Julier, Vinattieri,
  Colocci, Lefebvre, Gil, Scalbert, Tran, Karlicek, and Lascaray}}]{Julier}
\bibinfo{author}{\bibfnamefont{M.}~\bibnamefont{Julier}},
  \bibinfo{author}{\bibfnamefont{A.}~\bibnamefont{Vinattieri}},
  \bibinfo{author}{\bibfnamefont{M.}~\bibnamefont{Colocci}},
  \bibinfo{author}{\bibfnamefont{P.}~\bibnamefont{Lefebvre}},
  \bibinfo{author}{\bibfnamefont{B.}~\bibnamefont{Gil}},
  \bibinfo{author}{\bibfnamefont{D.}~\bibnamefont{Scalbert}},
  \bibinfo{author}{\bibfnamefont{C.}~\bibnamefont{Tran}},
  \bibinfo{author}{\bibfnamefont{R.}~\bibnamefont{Karlicek}}, \bibnamefont{and}
  \bibinfo{author}{\bibfnamefont{J.}~\bibnamefont{Lascaray}},
  \bibinfo{journal}{Phys. Stat. Sol. (b)} \textbf{\bibinfo{volume}{216}},
  \bibinfo{pages}{341} (\bibinfo{year}{1999}).

\bibitem[{\citenamefont{Tackeuchi et~al.}(2006)\citenamefont{Tackeuchi, Otake,
  Ogawa, Ushiyama, Fujita, Takano, and Akinaga}}]{Tackeuchi_2006}
\bibinfo{author}{\bibfnamefont{A.}~\bibnamefont{Tackeuchi}},
  \bibinfo{author}{\bibfnamefont{H.}~\bibnamefont{Otake}},
  \bibinfo{author}{\bibfnamefont{Y.}~\bibnamefont{Ogawa}},
  \bibinfo{author}{\bibfnamefont{T.}~\bibnamefont{Ushiyama}},
  \bibinfo{author}{\bibfnamefont{T.}~\bibnamefont{Fujita}},
  \bibinfo{author}{\bibfnamefont{F.}~\bibnamefont{Takano}}, \bibnamefont{and}
  \bibinfo{author}{\bibfnamefont{H.}~\bibnamefont{Akinaga}},
  \bibinfo{journal}{Appl. Phys. Lett.} \textbf{\bibinfo{volume}{88}},
  \bibinfo{pages}{162114} (\bibinfo{year}{2006}).

\bibitem[{\citenamefont{Paillard et~al.}(2001)\citenamefont{Paillard, Marie,
  Renucci, Amand, Jbeli, and G\'erard}}]{Paillard}
\bibinfo{author}{\bibfnamefont{M.}~\bibnamefont{Paillard}},
  \bibinfo{author}{\bibfnamefont{X.}~\bibnamefont{Marie}},
  \bibinfo{author}{\bibfnamefont{P.}~\bibnamefont{Renucci}},
  \bibinfo{author}{\bibfnamefont{T.}~\bibnamefont{Amand}},
  \bibinfo{author}{\bibfnamefont{A.}~\bibnamefont{Jbeli}}, \bibnamefont{and}
  \bibinfo{author}{\bibfnamefont{J.~M.}~\bibnamefont{G\'erard}},
  \bibinfo{journal}{Phys. Rev. Lett.} \textbf{\bibinfo{volume}{86}},
  \bibinfo{pages}{1634} (\bibinfo{year}{2001}).

\bibitem[{\citenamefont{Tsitsishvili et~al.}(2002)\citenamefont{Tsitsishvili,
  Baltz, and Kalt}}]{Kalt}
\bibinfo{author}{\bibfnamefont{E.}~\bibnamefont{Tsitsishvili}},
  \bibinfo{author}{\bibfnamefont{R.~v.}~\bibnamefont{Baltz}}, \bibnamefont{and}
  \bibinfo{author}{\bibfnamefont{H.}~\bibnamefont{Kalt}},
  \bibinfo{journal}{Phys. Rev. B} \textbf{\bibinfo{volume}{66}},
  \bibinfo{pages}{161405(R)} (\bibinfo{year}{2002}).

\bibitem[{\citenamefont{Flissikowski et~al.}(2003)\citenamefont{Flissikowski,
  Akimov, Hundt, and Henneberger}}]{Henneberger}
\bibinfo{author}{\bibfnamefont{T.}~\bibnamefont{Flissikowski}},
  \bibinfo{author}{\bibfnamefont{I.~A.}~\bibnamefont{Akimov}},
  \bibinfo{author}{\bibfnamefont{A.}~\bibnamefont{Hundt}}, \bibnamefont{and}
  \bibinfo{author}{\bibfnamefont{F.}~\bibnamefont{Henneberger}},
  \bibinfo{journal}{Phys. Rev. B} \textbf{\bibinfo{volume}{68}},
  \bibinfo{pages}{161309(R)} (\bibinfo{year}{2003}).

\bibitem[{\citenamefont{Martinez-Guerrero
  et~al.}(2000)\citenamefont{Martinez-Guerrero, Adelmann, Chabuel, Simon,
  Pelekanos, Mula, Daudin, Feuillet, and Mariette}}]{Mariette}
\bibinfo{author}{\bibfnamefont{E.}~\bibnamefont{Martinez-Guerrero}},
  \bibinfo{author}{\bibfnamefont{C.}~\bibnamefont{Adelmann}},
  \bibinfo{author}{\bibfnamefont{F.}~\bibnamefont{Chabuel}},
  \bibinfo{author}{\bibfnamefont{J.}~\bibnamefont{Simon}},
  \bibinfo{author}{\bibfnamefont{N.}~\bibnamefont{Pelekanos}},
  \bibinfo{author}{\bibfnamefont{G.}~\bibnamefont{Mula}},
  \bibinfo{author}{\bibfnamefont{B.}~\bibnamefont{Daudin}},
  \bibinfo{author}{\bibfnamefont{G.}~\bibnamefont{Feuillet}}, \bibnamefont{and}
  \bibinfo{author}{\bibfnamefont{H.}~\bibnamefont{Mariette}},
  \bibinfo{journal}{Appl. Phys. Lett.} \textbf{\bibinfo{volume}{77}},
  \bibinfo{pages}{809} (\bibinfo{year}{2000}).

\bibitem[{Pui()}]{Puissance}
\bibinfo{note}{Similar effects were observed with excitation powers in the
  range 0.5 - 5 mW}.

\bibitem[{\citenamefont{Heitz et~al.}(1997)\citenamefont{Heitz, Veit,
  Ledentsov, Hoffmann, Bimberg, Ustinov, Kopev, and Alferov}}]{Heitz}
\bibinfo{author}{\bibfnamefont{R.}~\bibnamefont{Heitz}},
  \bibinfo{author}{\bibfnamefont{M.}~\bibnamefont{Veit}},
  \bibinfo{author}{\bibfnamefont{N.~N.}~\bibnamefont{Ledentsov}},
  \bibinfo{author}{\bibfnamefont{A.}~\bibnamefont{Hoffmann}},
  \bibinfo{author}{\bibfnamefont{D.}~\bibnamefont{Bimberg}},
  \bibinfo{author}{\bibfnamefont{V.~M.}~\bibnamefont{Ustinov}},
  \bibinfo{author}{\bibfnamefont{P.~S.}~\bibnamefont{Kopev}}, \bibnamefont{and}
  \bibinfo{author}{\bibfnamefont{Z.~I.}~\bibnamefont{Alferov}},
  \bibinfo{journal}{Phys. Rev. B} \textbf{\bibinfo{volume}{56}},
  \bibinfo{pages}{10435} (\bibinfo{year}{1997}).

\bibitem[{Int()}]{Integration}
\bibinfo{note}{The time integration has been performed for t$>$50 ps to avoid
  the backscattered laser light from the sample surface}.

\bibitem[{\citenamefont{Bonnot et~al.}(1974)\citenamefont{Bonnot, Planel, and
  \`a~la Guillaume}}]{Planel}
\bibinfo{author}{\bibfnamefont{A.}~\bibnamefont{Bonnot}},
  \bibinfo{author}{\bibfnamefont{R.}~\bibnamefont{Planel}}, \bibnamefont{and}
  \bibinfo{author}{\bibfnamefont{C.~B.} \bibnamefont{\`a~la Guillaume}},
  \bibinfo{journal}{Phys. Rev. B} \textbf{\bibinfo{volume}{9}},
  \bibinfo{pages}{690} (\bibinfo{year}{1974}).

\bibitem[{\citenamefont{Meier and Zakharchenya}(1984)}]{Orientation}
\bibinfo{author}{\bibfnamefont{F.}~\bibnamefont{Meier}} \bibnamefont{and}
  \bibinfo{author}{\bibfnamefont{B.}~\bibnamefont{Zakharchenya}},
  \emph{\bibinfo{title}{Optical Orientation}} (\bibinfo{publisher}{North
  Holland, Amsterdam}, \bibinfo{year}{1984}).

\bibitem[{\citenamefont{Ivchenko}(1995)}]{Ivchenko}
\bibinfo{author}{\bibfnamefont{E.}~\bibnamefont{Ivchenko}},
  \bibinfo{journal}{Pure \& Appl. Chem.} \textbf{\bibinfo{volume}{67}},
  \bibinfo{pages}{463} (\bibinfo{year}{1995}).

\bibitem[{\citenamefont{Winkelnkemper et~al.}(2007)\citenamefont{Winkelnkemper,
  Seguin, Rodt, Schliwa, Reissmann, Strittmatter, Hoffmann, and
  Bimberg}}]{Bimberg}
\bibinfo{author}{\bibfnamefont{M.}~\bibnamefont{Winkelnkemper}},
  \bibinfo{author}{\bibfnamefont{R.}~\bibnamefont{Seguin}},
  \bibinfo{author}{\bibfnamefont{S.}~\bibnamefont{Rodt}},
  \bibinfo{author}{\bibfnamefont{A.}~\bibnamefont{Schliwa}},
  \bibinfo{author}{\bibfnamefont{L.}~\bibnamefont{Reissmann}},
  \bibinfo{author}{\bibfnamefont{A.}~\bibnamefont{Strittmatter}},
  \bibinfo{author}{\bibfnamefont{A.}~\bibnamefont{Hoffmann}}, \bibnamefont{and}
  \bibinfo{author}{\bibfnamefont{D.}~\bibnamefont{Bimberg}},
  \bibinfo{journal}{J. Appl. Phys.} \textbf{\bibinfo{volume}{101}},
  \bibinfo{pages}{113708} (\bibinfo{year}{2007}).

\bibitem[{Ang()}]{Angle}
\bibinfo{note}{We checked this for any crystallographic orientation for the
  detection}.

\bibitem[{\citenamefont{Simon et~al.}(2003)\citenamefont{Simon, Pelekanos,
  Adelmann, Martinez-Guerrero, Andr\'e, Daudin, Dang, and Mariette}}]{Simon}
\bibinfo{author}{\bibfnamefont{J.}~\bibnamefont{Simon}},
  \bibinfo{author}{\bibfnamefont{N.~T.}~\bibnamefont{Pelekanos}},
  \bibinfo{author}{\bibfnamefont{C.}~\bibnamefont{Adelmann}},
  \bibinfo{author}{\bibfnamefont{E.}~\bibnamefont{Martinez-Guerrero}},
  \bibinfo{author}{\bibfnamefont{R.}~\bibnamefont{Andr\'e}},
  \bibinfo{author}{\bibfnamefont{B.}~\bibnamefont{Daudin}},
  \bibinfo{author}{\bibfnamefont{L.~S.}~\bibnamefont{Dang}}, \bibnamefont{and}
  \bibinfo{author}{\bibfnamefont{H.}~\bibnamefont{Mariette}},
  \bibinfo{journal}{Phys. Rev. B} \textbf{\bibinfo{volume}{68}},
  \bibinfo{pages}{035312} (\bibinfo{year}{2003}).

\bibitem[{\citenamefont{Bretagnon et~al.}(2006)\citenamefont{Bretagnon,
  Lefebvre, Valvin, Bardoux, Guillet, Taliercio, Gil, Grandjean, Semond,
  Damilano et~al.}}]{Grandjean}
\bibinfo{author}{\bibfnamefont{T.}~\bibnamefont{Bretagnon}},
  \bibinfo{author}{\bibfnamefont{P.}~\bibnamefont{Lefebvre}},
  \bibinfo{author}{\bibfnamefont{P.}~\bibnamefont{Valvin}},
  \bibinfo{author}{\bibfnamefont{R.}~\bibnamefont{Bardoux}},
  \bibinfo{author}{\bibfnamefont{T.}~\bibnamefont{Guillet}},
  \bibinfo{author}{\bibfnamefont{T.}~\bibnamefont{Taliercio}},
  \bibinfo{author}{\bibfnamefont{B.}~\bibnamefont{Gil}},
  \bibinfo{author}{\bibfnamefont{N.}~\bibnamefont{Grandjean}},
  \bibinfo{author}{\bibfnamefont{F.}~\bibnamefont{Semond}},
  \bibinfo{author}{\bibfnamefont{B.}~\bibnamefont{Damilano}},
  \bibnamefont{et~al.}, \bibinfo{journal}{Phys. Rev. B}
  \textbf{\bibinfo{volume}{73}}, \bibinfo{pages}{113304}
  (\bibinfo{year}{2006}).

\bibitem[{\citenamefont{Lenihan et~al.}(2002)\citenamefont{Lenihan, GurudevDutt, 
  Steel, Ghosh, and Bhattacharya}}]{Steel}
\bibinfo{author}{\bibfnamefont{A.~S.} \bibnamefont{Lenihan}},
  \bibinfo{author}{\bibfnamefont{M.~V.} \bibnamefont{GurudevDutt}},
  \bibinfo{author}{\bibfnamefont{D.~G.} \bibnamefont{Steel}},
  \bibinfo{author}{\bibfnamefont{S.}~\bibnamefont{Ghosh}}, \bibnamefont{and}
  \bibinfo{author}{\bibfnamefont{P.~K.} \bibnamefont{Bhattacharya}},
  \bibinfo{journal}{Phys. Rev. Lett.} \textbf{\bibinfo{volume}{88}},
  \bibinfo{pages}{223601} (\bibinfo{year}{2002}).

\bibitem[{\citenamefont{Kusrayev et~al.}(2005)\citenamefont{Kusrayev, Koudinov,
  Zakharchenya, Lee, Furdyna, and Dobrowolska}}]{Kusrayev}
\bibinfo{author}{\bibfnamefont{Y.~G.} \bibnamefont{Kusrayev}},
  \bibinfo{author}{\bibfnamefont{A.~V.} \bibnamefont{Koudinov}},
  \bibinfo{author}{\bibfnamefont{B.~P.} \bibnamefont{Zakharchenya}},
  \bibinfo{author}{\bibfnamefont{S.}~\bibnamefont{Lee}},
  \bibinfo{author}{\bibfnamefont{J.~K.} \bibnamefont{Furdyna}},
  \bibnamefont{and}
  \bibinfo{author}{\bibfnamefont{M.}~\bibnamefont{Dobrowolska}},
  \bibinfo{journal}{Phys. Rev. B} \textbf{\bibinfo{volume}{72}},
  \bibinfo{pages}{155301} (\bibinfo{year}{2005}).

\bibitem[{\citenamefont{Vurgaftman and Meyer}(2003)}]{Varshni}
\bibinfo{author}{\bibfnamefont{I.}~\bibnamefont{Vurgaftman}} \bibnamefont{and}
  \bibinfo{author}{\bibfnamefont{J.}~\bibnamefont{Meyer}}, \bibinfo{journal}{J.
  Appl. Phys.} \textbf{\bibinfo{volume}{94}}, \bibinfo{pages}{3675}
  (\bibinfo{year}{2003}).

\bibitem[{\citenamefont{Ramvall et~al.}(1998)\citenamefont{Ramvall, Tanaka,
  Nomura, Riblet, and Aoyagi}}]{Ramvall}
\bibinfo{author}{\bibfnamefont{P.}~\bibnamefont{Ramvall}},
  \bibinfo{author}{\bibfnamefont{S.}~\bibnamefont{Tanaka}},
  \bibinfo{author}{\bibfnamefont{S.}~\bibnamefont{Nomura}},
  \bibinfo{author}{\bibfnamefont{P.}~\bibnamefont{Riblet}}, \bibnamefont{and}
  \bibinfo{author}{\bibfnamefont{Y.}~\bibnamefont{Aoyagi}},
  \bibinfo{journal}{Appl. Phys. Lett.} \textbf{\bibinfo{volume}{73}},
  \bibinfo{pages}{1104} (\bibinfo{year}{1998}).

\bibitem[{\citenamefont{Xia and Wei}(2006)}]{Xia}
\bibinfo{author}{\bibfnamefont{C.}~\bibnamefont{Xia}} \bibnamefont{and}
  \bibinfo{author}{\bibfnamefont{S.}~\bibnamefont{Wei}},
  \bibinfo{journal}{Microelectronics Journal} \textbf{\bibinfo{volume}{37}},
  \bibinfo{pages}{1408} (\bibinfo{year}{2006}).

\bibitem[{\citenamefont{Dzhioev et~al.}(1998)\citenamefont{Dzhioev,
  Zakharchenya, Ivchenko, Korenev, Kusraev, Ledentsov, Ustinov, Zhukov, and
  A.F.Tsatsul'nikov}}]{Dzhioev}
\bibinfo{author}{\bibfnamefont{R.}~\bibnamefont{Dzhioev}},
  \bibinfo{author}{\bibfnamefont{B.}~\bibnamefont{Zakharchenya}},
  \bibinfo{author}{\bibfnamefont{E.}~\bibnamefont{Ivchenko}},
  \bibinfo{author}{\bibfnamefont{V.}~\bibnamefont{Korenev}},
  \bibinfo{author}{\bibfnamefont{Y.}~\bibnamefont{Kusraev}},
  \bibinfo{author}{\bibfnamefont{N.}~\bibnamefont{Ledentsov}},
  \bibinfo{author}{\bibfnamefont{V.}~\bibnamefont{Ustinov}},
  \bibinfo{author}{\bibfnamefont{A.}~\bibnamefont{Zhukov}}, \bibnamefont{and}
  \bibinfo{author}{\bibnamefont{A.F.Tsatsul'nikov}}, \bibinfo{journal}{Phys.
  Solid State} \textbf{\bibinfo{volume}{40}}, \bibinfo{pages}{790}
  (\bibinfo{year}{1998}).

\end{thebibliography}
\end{document}